\documentclass[aps,preprint,superscriptaddress,floatfix,nofootinbib,onecolumn,longbibliography]{revtex4-1}

\usepackage{bm}
\usepackage{bbm}
\usepackage{bbold}
\usepackage{bbm}
\usepackage[pdftex]{graphicx}
\usepackage{latexsym,amsmath,verbatim,amssymb,txfonts}
\usepackage{color}
\usepackage{rotating}
\usepackage{verbatim}
\usepackage{multirow}
\usepackage[english]{babel}
\usepackage{comment}

\begin{document}
\def\be{\begin{equation}}
\def\ee{\end{equation}}

\def\bc{\begin{center}}
\def\ec{\end{center}}
\def\bea{\begin{eqnarray}}
\def\eea{\end{eqnarray}}
\newcommand{\avg}[1]{\langle{#1}\rangle}
\newcommand{\Avg}[1]{\left\langle{#1}\right\rangle}
\newcommand{\mi}{\mathrm{i}} %% roman "i"
\newcommand{\di}{i}    
\def\ie{\textit{i.e.}}
\def\etal{\textit{et al.}}
\def\m{\vec{m}}
\def\G{\mathcal{G}}

\title{Grand canonical ensembles of sparse  networks and Bayesian inference}

\author{Ginestra Bianconi}
\affiliation{School of Mathematical Sciences, Queen Mary University of London, London, E1 4NS, United Kingdom}
\affiliation{The Alan Turing Institute, The British Library, London NW1 2DB, United Kingdom}
\email{ginestra.bianconi@gmail.com}

\begin{abstract}
Maximum entropy  network ensembles have been very successful in modelling sparse network topologies and in solving challenging inference problems. However the sparse maximum entropy network models proposed so far have fixed number of nodes and are typically not exchangeable. Here we consider hierarchical models for  exchangeable networks in the sparse limit, i.e. with the total number of links scaling linearly with the total number of nodes. The approach is grand canonical, i.e. the number of nodes of the network is not fixed a priori: it is finite but can be arbitrarily large. In this way the grand canonical network ensembles  circumvent the difficulties in treating  infinite sparse exchangeable networks which according to  the  Aldous-Hoover theorem must vanish.
The approach can treat networks with given degree distribution or networks with given distribution of latent variables.
When only a subgraph induced by a subset  of nodes is known,  this model allows a  Bayesian estimation of the network size and the degree sequence (or the sequence of latent variables) of the entire network which can be used for network reconstruction.
\end{abstract}

\maketitle

\section{Introduction}
Networks \cite{Book_Laszlo,Book_Newman} have the ability to capture the topology of complex systems ranging from the brain to financial networks. Network models are key to have reliable unbiased null models of the network and to explain emergent phenomena of network evolution. Network model can be classified in two major classes: equilibrium maximum entropy models  \cite{anand2009entropy,park2004statistical,bianconi2021information,cimini2019statistical,krioukov2010hyperbolic,orsini2015quantifying,peixoto2012entropy,radicchi2020classical,pessoa2021entropic,kim2012constructing,del2010efficient,coolen2017generating,bassler2015exact} and growing network models \cite{BA,Book_Laszlo,Doro_book,kharel2021degree}.
While growing network models have a number of nodes that increases in time, maximum entropy models are used so far only for  treating networks of a given number of nodes $N$. In this paper we are  interested in extending the realm of maximum entropy network models to networks of varying network size $N$.

Maximum entropy network ensembles are the least biased ensembles satisfying a given set of constraints. As such maximum entropy ensembles are widely used as null models and for network reconstruction starting from features associated  to the nodes of the network. Given the profound relation between information theory and statistical mechanics \cite{jaynes1957information,huang2009introduction}, maximum entropy  network ensembles can be distinguished between microcanonical ensembles and canonical ensembles \cite{anand2009entropy,anand2010gibbs,bianconi2008entropies} similarly to the analogous distinction traditionally introduced in statistical mechanics for  ensembles of particles. Microcanonical  network ensembles are ensembles of networks of  $N$ nodes satisfying some hard constraints (such as the total number of links, or the given degree sequence). Canonical network ensembles instead are ensemble of networks of $N$ nodes satisfying some soft constraints, (such as the  expected total number of links or the expected degree sequence). The canonical ensembles with expected degree sequence can be also formulated as latent variable models  where  the latent variables can be associated to the nodes \cite{capocci,bianconi2021information}.

Maximum entropy models have been very successful in solving  challenging inference models \cite{PNAS,airoldi2008mixed,cimini2019statistical,orsini2015quantifying,ghavasieh2020statistical}, however they have the limitation that they only treat networks with a given fixed number of nodes $N$. Indeed in a number of scenarios, the number of nodes might not be fixed or might not be known. In this context an important problem is  to compare networks of different network sizes. For instance in brain imaging one might choose a finer grid  or a coarser grid of brain regions and an outstanding problem in machine learning is how to build  neural networks that can generalize well when tested on network data with  different 
network size than the network data in the training set \cite{bevilacqua2021size,Riberio}.

In order to have network ensembles that can treat networks of different size, here  we introduce the grand canonical network ensembles in which the number of nodes  can vary.
A well-defined grand-canonical network ensemble  necessarily needs to be exchangeable \cite{deFinetti}, i.e. needs to be invariant under permutation of labels of the nodes of the network, so that removing or adding a node has an effect that is independent of   the particular choice of the node added or removed. 

The research on exchangeable networks is currently very vibrant. The graphon model \cite{lovasz2012large} is the most well established exchangeable network model. However this model is dense, i.e. the number of links scales quadratically with the number of nodes while the vast majority of the network data is sparse with a total number of links scaling linearly with the network size. In other words most of the real world networks have constant average degree. However popular models for sparse networks such as the configuration model \cite{chung2002average} and the exponential random graphs \cite{park2004statistical} are not exchangeable. In fact these models treat networks of labelled nodes with given degree or with given expected degree sequence. Therefore the network ensemble is not invariant under permutation of the node labels, except if all the degrees of all the expected degrees of the network are the same (for a more diffused discussion of why these networks are not exchangeable see discussion in Ref.\cite{bianconi2022exchangeable}). Several works have been proposed  exchangeable network models in the  when the average degree of the network diverges sublinearly with the network size  \cite{caron2017sparse,borgs2016sparse,veitch2015class,veitch2019sampling,borgs2015private,borgs2018revealing}.  In Ref. \cite{bianconi2022exchangeable} a framework able to model  sparse exchangeable networks in the limit of constant degree, has been proposed. The model is very general  and has been extended to treat generalized network structures including multiplex networks \cite{bianconi2018multilayer} and simplicial complexes \cite{bianconi2021higher}. However the model is well defined only for finite networks of large but finite number of nodes $N$ as exchangeable sparse networks need to obey the Aldous-Hoover theorem \cite{Aldous,Hoover} according to which infinite sparse exchangeable networks must vanish. An alternative strategy for formulating exchangeable ensembles is to consider ensembles of unlabelled networks for which several results are already available \cite{krioukov2022}.  

Here we build on the recently proposed exchangeable sparse network ensembles \cite{bianconi2022exchangeable} to formulate hierarchical grand-canonical ensembles of sparse networks.
The proposed grand-canonical ensembles are hierarchical models \cite{airoldi2008mixed,peixoto2014hierarchical} with variable number of nodes $N$ and with given degree distribution or alternatively given latent variable distributions.
The grand canonical approach provides a way to circumvent the limitations imposed by the Aldous-Hoover theorem because in this framework one considers a mixture of network ensembles with finite but unspecified and arbitrary large network sizes.
In this paper  we define the grand-canonical ensembles and we characterize them with statistical mechanics methods, evaluating their entropy, the marginal probability of a link and  proposing generative algorithms to sample networks from these ensembles.
[Note that the proposed grand canonical ensembles differ from the ensembles proposed in Refs. \cite{gabrielli2019grand,straka2017grand}, as in our case we consider networks with undetermined number of nodes, while in Refs.\cite{gabrielli2019grand,straka2017grand} is the total sum of weights of weighted networks that is allowed to vary. From the statistical mechanics perspective our approach is fully classical while in Refs. \cite{gabrielli2019grand,straka2017grand} networks ensembles are treated as quantum mechanical ensembles where the particles are associated to the links of the network and the adjacency matrix elements play the role of occupation numbers.]

Finally, we use the gran-canonical network ensembles to solve an inference problem. We  consider a scenario in which the entire network has an unknown number of nodes, and we have only access to a subgraph induced by a subset of its nodes. In this 
hypothesis we use the grand-canonical network models to perform a Bayesian estimation of the true parameters of the network model (given by the network size and the degree sequence or the sequence of latent variables). This a posteriori estimate of the parameters can then be used to reconstruct the unknown part of the network.

\section{The  grand canonical network ensemble with given degree distribution}
\begin{figure*}[htbp!]
		\centering
	\includegraphics[width=0.95\textwidth]{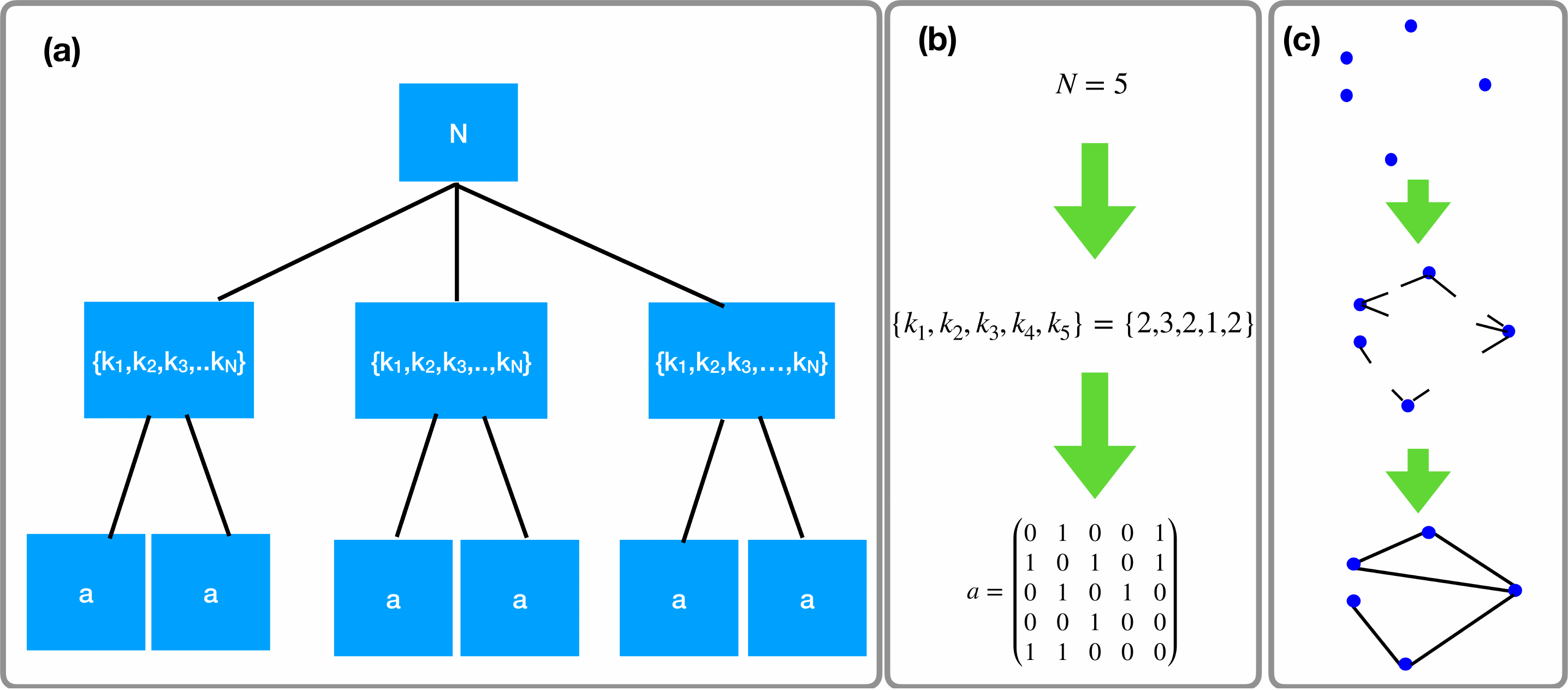}
	\caption{Schematic representation of the hierarchical grand canonical  ensemble of exchangeable sparse simple networks.  The proposed ensemble is a hierarchical model of networks in which first the total number of nodes $N$ is drawn from a $P(N)=\pi(N)$ distribution, then a given degree sequence  ${\bf k}=\{k_1,k_2,\ldots k_N\}$ is drawn from the distribution $P({\bf k}|N)$ among all the degree sequence with the total number of nodes $N$; finally a  network $G$ with adjacency matrix ${\bf a}$ drawn from the distribution $P(G|{\bf k},N)$ among all the networks with a given total number of nodes $N$ and degree sequence ${\bf k}$. Panel (a) describes the hierarchical nature of the model, panel (b) provide an example of subsequent draw of the total number of nodes, the degree sequence and the adjacency matrix of the network, panel (c) is a visualization of the construction of a network according to the proposed model.  }
	\label{fig:1}
	\end{figure*}

We consider the hierarchical grand canonical  ensemble of exchangeable sparse simple networks where we associate to every network $G=(V,E)$ with $N=|V|>N_0$ nodes the probability 
%\cite{bianconi2009entropy}\cite{anand2009entropy}
\begin{equation}
    \mathbb{P}(G)=P(N)P({\bf k}|N)P({G}|{\bf k},N)
\end{equation}
where $P(N)$ indicates the probability that the network $G$ has $N$ nodes, $P({\bf k}|N)$ indicates the conditional probability that the network has degree sequence ${\bf k}$ given that the network has $N$ nodes, and $P({G}|{\bf k},N)$ indicates the probability of the network $G$ with adjacency matrix ${\bf a}$ given that the network has $N$ nodes and degree sequence ${\bf k}$ (see Figure \ref{fig:1} for a schematic representation of the model).
To be specific we consider the following model giving rise to the hierarchical grand canonical  ensemble of exchangeable simple models:
\begin{itemize}
\item[(1)]{\em Drawing the total number of nodes $N$ of the network.}
Let us discuss suitable choices for the distribution of the number of nodes $N$ with $N$ greater or equal than some minimum number of nodes $N_0$. We  indicate the distribution $P(N)$ as 
\bea
P(N)&=&\pi(N),\quad \mbox{for} \  N\geq N_0.
\eea
While a statistical mechanics approach would suggest to take a distribution $\pi(N)$ with a well defined mean value (such as the exponential distribution)
\bea
\pi(N)=Ce^{-\mu N} \quad\mbox{for } N\geq N_0,
\label{p1}
\eea
where $C$ is a  normalization constant and $\mu>0$,
in the context of network science it might actually be relevant to consider also broad distributions $\pi(N)$ such as power-law distributions 
\bea
\pi(N)=D N^{-\nu} \quad\mbox{for } N\geq N_0,
\label{p2}
\eea
where $D$ is a normalization constant and  $\nu>1$.
\item[(2)]{\em Drawing the degree sequence of the network.}
In order to obtain a sparse exchangeable network ensemble with given degree distribution $p(k)$ having finite average degree $\avg{k}$,  minimum allowed degree $\hat{m}$ and maximum allowed degree $K$ we consider the following expression for the probability of a given degree sequence given the total number of nodes 
\bea
P({\bf k}|N)&=&\prod_{i=1}^N \left[p(k_i)\hat{\theta}(K-k_i)\theta(k_i-\hat{m})\right]\delta\left(\sum_{i=1}^N k_i,\avg{k}N\right),
\eea
where  $\hat{\theta}(x)$ indicates the Heaviside function $\hat{\theta}(x)=1$ if $x\geq 0$ and $\hat{\theta}(x)=0$ otherwise  and where we used the notation $\avg{k}=\sum_kkp(k)$. In the following we will indicate with $L$ the total number of links of the network given by $L=\avg{k}N/2$. 
Note that $P({\bf k}|N)$ is independent of the labels of the nodes, i.e. all the degree sequences that can be  obtained by a permutation of the node labels  of a given degree sequence have the same probability $P({\bf k}|N)$.
\item[(3)]{\em Drawing the adjacency matrix of the network.}
The probability of a network $G$ with adjacency matrix ${\bf a}$ given the total number of nodes $N$ of the network and the degree sequence ${\bf k}$ is chosen in the least biased way by drawing the network from a uniform distribution, i.e. the conditional probability $P({G}|{\bf k},N)$ is equivalent to the probability of a network in the microcanonical ensemble.
Therefore, by indicating with $\mathcal{N}({\bf k}|N)$ the total number of networks with $N$ nodes and degree sequence ${\bf k}$ and with $\Sigma_N({\bf k})=\ln \mathcal{N}({\bf k}|N)$ the entropy of the ensemble we can express $P({G}|{\bf k},N)$ as
\bea
P({G}|{\bf k},N)&=&\frac{1}{\mathcal{N}({\bf k}|N)}=e^{-\Sigma_N({\bf k})}
\eea
Note that for sparse networks of $N\geq N_0$ nodes  the entropy $\Sigma_N({\bf k})$  obeys the Bender-Canfield formula as long as the network has a structural cutoff $K_S$, i.e. as long as $k_i\ll K_S=\sqrt{\avg{k}N_0}$  \cite{bender,bianconi2008entropies,anand2009entropy,anand2010gibbs}
\bea
{\Sigma}_N({\bf k})= \ln \left(\frac{(2L)!!}{\prod_{i=1}^Nk_i!}\right)+o(N)
\label{S}
\eea
 where in Eq. (\ref{S}) we indicate with ${\bf k}=\{k_1,k_2,\ldots, k_N\}$ the degree sequence with    $k_i$, the degree of node $i$, given by $k_i=\sum_{j=1}^Na_{ij}$.
\end{itemize}

It follows that the hierarchical grand canonical  ensemble for exchangeable sparse networks can be cast into an Hamiltonian ensemble with probability $\mathbb{P}(G)$ given by 
\bea
\mathbb{P}(G)&=&\frac{1}{Z}e^{-H(G)}\delta\left(\avg{k}N/2,\sum_{i<j}a_{ij}\right)\hat{\theta}\left(K-\max_{i=1}^Nk_i\right)\hat{\theta}\left(\min_{i=1}^Nk_i-\hat{m}\right),\label{PG}
\eea
with Hamiltonian $H(G)$ given by 
\bea
H(G)&=&-\ln \pi(N)-\sum_{i=1}^{N}\ln \left[p(k_i)k_i! \delta\left(k_i,\sum_{j=1}^Na_{ij}\right)\right]+\ln ((\avg{k}N) !!).
\label{Hamiltonian}
\eea
This Hamiltonian is global and is invariant under permutation of the node labels,
therefore this hierarchical grand canonical ensemble is exchangeable.
Indeed we have that the probability of a network $\mathbb{P}(G)$ given by Eq. (\ref{PG}) obeys
\bea
\mathbb{P}(G)=\mathbb{P}(\tilde{G})
\eea
where $\tilde{G}$ is any network  obtained from network $G$ under a generic  permutation $\sigma$ of the labels of the nodes.
Moreover we note that for $\pi(N)=\delta(N,\bar{N})$, i.e. when the network size is fixed this model reduces to the exchangeable model for sparse network ensemble proposed in Ref.\cite{bianconi2022exchangeable}.

\section{The  grand canonical network ensemble with given distribution of the latent variables}

The grand canonical formalism can also be easily extended to treat network models with latent variables $\bm\theta$ associated to the nodes of the network $G=(V,E)$. Note that here and in the following we assume that the latent variables take discrete values.
To this end we can consider the soft grand canonical hierarchical model  associating to each network with $N=|V|>N_0$ nodes, latent variables $\bm\theta$ and adjacency matrix ${\bf a}$ the probability 
\bea
P(G,\bm\theta,N)=P(N)P(\bm\theta|N)P({G}|\bm\theta, N)
\eea
with 
\bea
P(N)=\pi(N),
\eea
where $\pi(N)$ is an arbitrary prior on the number of nodes in the network defined for $N\geq N_0$. Typical examples of the distribution $\pi(N)$ are given by Eq. (\ref{p1}) and Eq. (\ref{p2}).
The probability of the latent variables is chosen to be exchangeable and given by 
\bea
P({\bm{\theta}}|N)=\prod_{i=1}^Np(\theta_i)
\eea
where $p(\theta_i)$ is the probability distribution of each latent variable. The distribution $p(\theta)$ can be chosen arbitrarily, as long as the expectation of $\theta$ is finite.
The probability of the network given the network size and the latent variables is chosen to be derived by a Bernoulli variable for each link, with probability of observing a link between node $i$ and node $j$ conditioned on the value of their latent variables  given by $p_N(\theta_i,\theta_j)$, i.e.
\bea 
P({G}|\bm\theta,N)=\prod_{i<j}p_N(\theta_i,\theta_j)^{a_{ij}}(1-p_N(\theta_i,\theta_j))^{1-a_{ij}}.
\eea
To be concrete we consider the following expression for the probability $p_N(\theta_i,\theta_j)$  which is the general expression of the marginal probability of a link in   canonical network ensembles (or equivalently exponential random graph models),
\bea
p_N(\theta_i,\theta_j)=\frac{\theta_i\theta_j/N}{1+\theta_i\theta_j/N}.
\label{ptheta2}
\eea
 The advantage of taking this expression for the probability $p_N(\theta_i,\theta_j)$ is that $p_N(\theta_i,\theta_j)$ is always smaller or equal to one  for every value of the latent variables. Therefore in this model we do not need  to impose a structural cutoff on the latent variables.
 In summary the grand canonical  network ensemble with given  latent variable distribution is a hierarchical network model  in which given the network size and latent variables the network is drawn according to a canonical ensemble of networks.
 In this ensemble the probability of a network $G$ can be written in Hamiltonian form as
\bea
\mathbb{P}(G)=\frac{1}{Z}e^{-H(G)}
\eea
with Hamiltonian $H(G)$ given by
\bea
H(G)=-\ln \pi(N)-\sum_{i=1}^Np_N(\theta_i)-\sum_{i<j}\left\{a_{ij}\ln p_N(\theta_i,\theta_j)+(1-a_{ij})\ln [1-p_N(\theta_i,\theta_j)]\right\}.
\eea
This Hamitonian is invariant under permutation of the node labels, therefore this model is exchangeable.
\section{The entropy of grand canonical ensembles}
In this paragraph we show  that the entropy $S$ \cite{bianconi2009entropy,anand2009entropy} of the two proposed grand canonical network ensembles, defined as
\bea
S=\sum_{G}\mathbb{P}(G)\ln \mathbb{P}(G),
\eea
can be decomposed into contributions that reflect the  uncertainty related to an increasing number of hierarchical levels of the model. In order to show this results we discuss separately the entropy of the two proposed grand canonical ensembles.  
\subsection{Entropy of the grand canonical ensemble with given degree distribution }
The entropy $S$ of the ensemble fixing the degree distribution can be decomposed into the entropy of the model at different levels of the hierarchy according to the following expression,
\bea
S=S_{\pi(N)}+\avg{S_{p(k)}}_{{\pi(N)}}+\Avg{\Sigma_N({\bf k})}_{\pi(N),p(k)}
\eea
where $S_{\pi(N)}$ is the entropy associated to the number of typical choices of the total number of nodes $N$, $\avg{S_{p(k)}}_{{\pi(N)}}$ is the entropy associated to the choice of the degree sequence averaged over the distribution $\pi(N)$ and $\Avg{\Sigma_N({\bf k})}_{\pi(N),p(k)}$ is the average of the Gibbs entropy \cite{anand2009entropy} of the networks with given degree sequence averaged over the distribution $\pi(N)$ and $P({\bf k}|N)$.
In other words we have 
\bea
S_{\pi(N)}&=&-\sum_{N>N_0}\pi(N)\ln \pi(N),\nonumber\\
\avg{S_{p(k)}}_{\pi(N)}&=&\sum_{N>N_0}\pi(N)\left[-N\sum_{k}p(k)\ln p(k)\right],\nonumber\\
\Avg{\Sigma_N({\bf k})}_{\pi(N),p(k)}&=&\sum_{N>N_0}\pi(N)\sum_{\bf k}P({\bf k}|N)\Sigma_N({\bf k}).
\eea 
\subsection{Entropy of the grand canonical ensemble with given latent variable distribution }
Similarly to the previous case, it is easy to show that the entropy of the ensemble fixing the  distribution of the latent variables can be decomposed into the entropy of the model at different levels of their hierarchy, according to the following expression
\bea
S=S_{\pi(N)}+\avg{S_{p(\theta)}}_{{\pi(N)}}+\Avg{S_{N}({\bm \theta})}_{\pi(N),p(\theta)},
\eea
where $S_{\pi(N)}$ is the entropy associated to the number of typical choices of the total number of nodes $N$, $\avg{S_{p(\theta)}}_{{\pi(N)}}$ is the entropy associated to the choice of the latent variable distribution averaged over the distribution $\pi(N)$ and $\Avg{S_N({\bm\theta})}_{\pi(N),p(k)}$ is the average of the Shannon  entropy  \cite{anand2009entropy} of the networks with given sequence of latent variables averaged over the distribution $\pi(N)$ and $P({\bm\theta}|N)$.
In other words we have 
\bea
S_{\pi(N)}&=&-\sum_{N>N_0}\pi(N)\ln \pi(N),\nonumber\\
\avg{S_{p(\theta)}}_{\pi(N)}&=&\sum_{N>N_0}\pi(N)\left[-N\sum_{\theta}p(\theta)\ln p(\theta)\right],\nonumber\\
\Avg{S_N({\bm\theta})}_{\pi(N),p(\theta)}&=&\sum_{N>N_0}\pi(N)\sum_{\bm\theta}P({\bm \theta}|N)S_N({\bm\theta}),
\eea 
where the Shannon entropy $S_N({\bm\theta})$ of the network given the sequence of latent variables and the network size $N$ can be expressed as
\bea
S_N({\bm\theta})=-\sum_{i<j}\left[p_N(\theta_i,\theta_j)\ln p_N(\theta_i,\theta_j)+(1-p_N(\theta_i,\theta_j))\ln (1-p_N(\theta_i,\theta_j))\right].
\eea
\section{Marginal probability of a link}
\subsection{The case of the  the grand canonical ensemble with given degree distribution }
The  grand canonical  ensemble of exchangeable sparse network ensembles is an ensemble in  which the total number of nodes is not specified. 
If we consider the networks of this ensemble having a given number of nodes $N$, the model reduces to the exchangeable sparse network ensemble proposed in Ref.\cite{bianconi2022exchangeable} whose marginal probability of a link $(i,j)$ is given by 
\bea
\tilde{p}_{ij}=\sum_{k}p(k)\sum_{k'}p(k')\frac{kk'}{\avg{k}N}.
\eea
Since the   grand-canonical  ensemble of sparse exchangeable networks with given degree distribution  can be interpreted as a mixture of the exchangeable sparse models proposed in Ref. \cite{bianconi2022exchangeable} with different size $N$, it is  immediate to show that  the marginal probability of a link between node $i$ and node $j$ in the grand canonical ensembles is given by the exchangeable expression, 
\bea
p_{ij}=\sum_{N>N_0}\pi(N)\sum_{k,k'}p(k)p(k')\frac{kk'}{\avg{k}N}=\sum_{N>N_0}\pi(N)\frac{\avg{k}}{N}.
\label{marginal}
\eea
%Therefore the probability $p_{ij}$ of a link between node $i$ and node $j$ is independent on $N$.
Moreover the probability that two nodes are connected given that they have degree $k$ and $k'$ is given by 
\bea
p_{ij|k_i=k,k_j=k'}=p({k,k'})=kk'\sum_{N>N_0}\frac{\pi(N)}{\avg{k}N}.
\eea
Finally the probability that two nodes are connected given that they have degree $k$ and $k'$ and the actual size of the network is $N$  is given by the uncorrelated network expression
\bea
p_{ij|k_i=k,k_j=k',N}=p_N(k,k')=\frac{kk'}{\avg{k}N}.
\label{conditioned}
\eea
From these expressions of the marginal probability of a link it is possible to appreciate how the hierarchical grand canonical ensemble of sparse exchangeable networks circumvents the difficulties arising form the Aldous-Hoover theorem without violating it. Indeed the marginal probability $p_N(k,k')$ of a link conditioned on the degrees of the two linked nodes and the  number of  nodes $N$ of the network vanishes in the limit $N\to\infty$, however if the number of nodes of the network is arbitrarily large but unknown the marginal probability of the link  remains finite (as both $p_{ij}$ and $p(k,k')$ are finite).
\begin{figure*}[htbp!]
		\centering
	\includegraphics[width=0.95\textwidth]{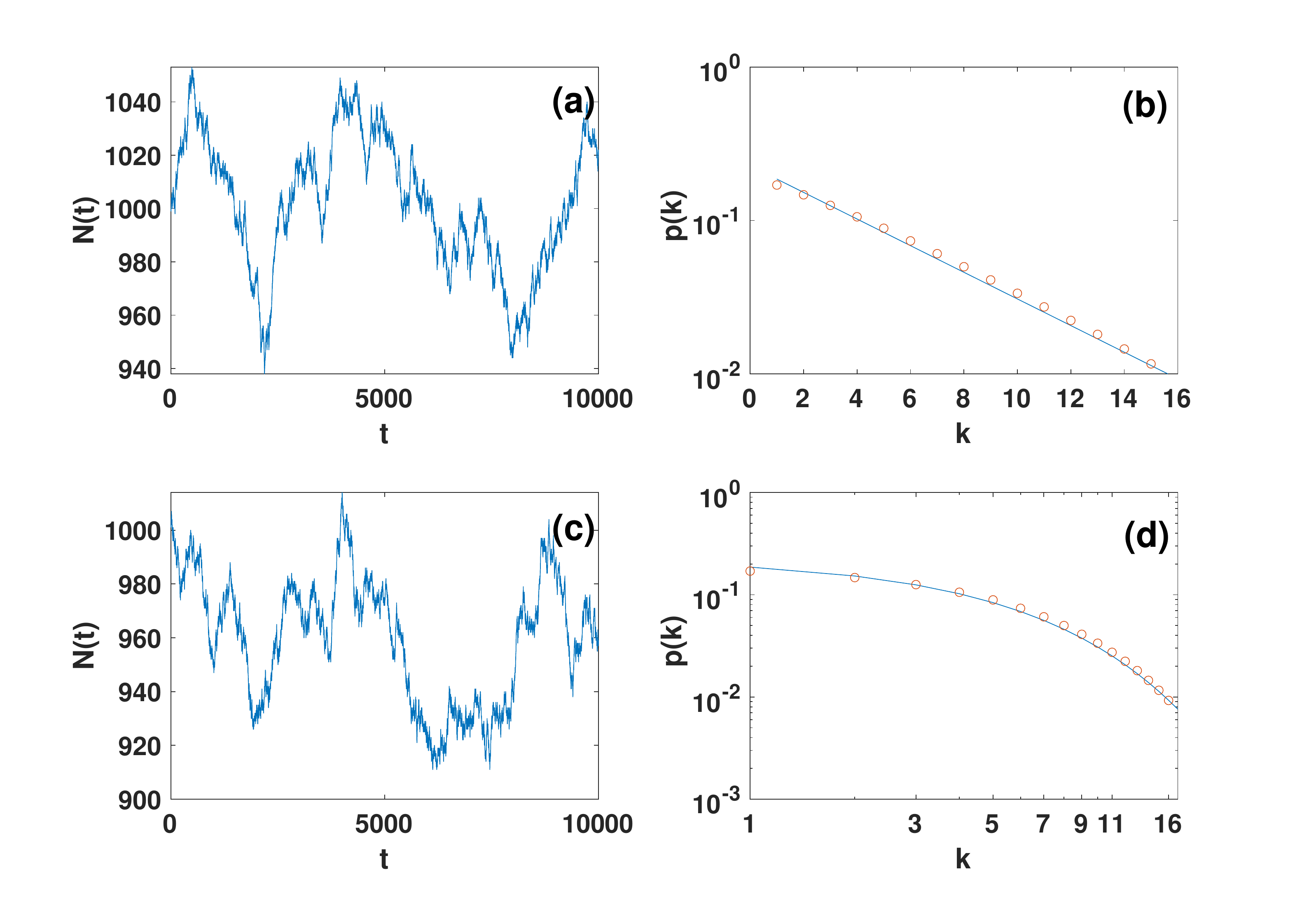}
	\caption{The number of nodes $N(t)$ at a function of time $t$ in the Metropolis-Hastings simulation of an exponential networks (panel a) and  networks with more general degree distribution (panel c) are shown together with the average degree distribution of the networks that is stable as the number of networks varies (symbols of panel (b) and (d)). The solid lines in panel(b) and panel (d) indicate the target degree distributions $p(k)=Ce^{-k/m}$ with $m=5$  (for panel (b)) and $p(k)=C(3+k)^{-\gamma}$ with $\gamma=3.4$ (for panel (d)).
The prior on the number of nodes is taken to be exponential $\pi(N)=Ce^{-N/\bar{N}}$ with $\bar{N}=1000$ with $N_0=500$ and $K=16$.}
	\label{fig:2}
	\end{figure*}
\subsection{The case of the  the grand canonical ensemble with given latent variable distribution }
For the grand canonical ensemble with given latent variable distribution $p(\theta)$ we have that the marginal probability of a link is given by 
\bea
p_{ij}=\sum_{N>N_0}\pi(N)\sum_{\theta,\theta'}p(\theta)p(\theta')p_N(\theta,\theta').
\eea
The probability of the link given the latent variable of the nodes is given by 
\bea
p(\theta,\theta')=\theta\theta'\sum_{N>N_0}\pi(N)\frac{1}{N+\theta\theta'};
\eea
the probability of a link given the network size and the latent variables is given by 
\bea
p_N(\theta,\theta')=\frac{\theta\theta'/N}{1+\theta\theta'/N}.
\eea
As we discussed in the case of the grand canonical ensemble  with given degree distribution also for the grand canonical ensemble with given latent variable distribution the grand canonical approach allows to circumvent the Aldous-Hoover theorem without violating it as the marginal probability of a link in an arbitrarily large network of unknown size is finite.
\section{Generating single instances of grand-canonical network ensembles}
In this section we describe two algorithms to generate single instances of the  proposed grand canonical ensembles. In particular we will discuss a Metropolis-Hastings ensemble to generate single instances of networks drawn from the grand canonical ensemble with given degree distribution and a Monte Carlo algorithm to generate single instances of networks drawn from the grand canonical ensemble with given distribution of latent variables.
\subsection{Metropolis-Hastings algorithm for the grand-canonical ensemble with given degree distribution} The grand-canonical   exchangeable ensemble of sparse networks can be obtained by  implementing a  Metropolis-Hastings algorithm using the network Hamiltonian given by Eq.(\ref{Hamiltonian}). 
\begin{itemize}
\item[(1)] Start with a network of $N$ nodes having exactly $L=\avg{k}N/2$ links and in which the minimum degree is greater of equal to $\hat{m}$ and the maximum degree is smaller or equal to $K$.
\item[(2)] Perform the Metropolis-Hastings algorithm for exchangeable sparse networks with $N$ nodes (defined below);
\item[(3)] 
Propose to change the number of nodes to $N'=N+1$ (addition of one node) or $N'=N-1$ (removal of one node) with equal probability and  accept the move with probability 
$\max\left(1,{\pi(N')}/{\pi(N)}\right)$ as long as $N'>N_0$. If the move is accepted change the number of nodes adding or removing a node, set the number of links to $L=\avg{k}N/2$ and ensure that each node has minimal degree at least $\hat{m}$ and maximum degree less than $K$. In particular if a node is added ensure it has at least $\hat{m}$ links by rewiring randomly the existing links of the networks and adding a number of links so that the total number of links is the integer that better approximates  $\avg{k}N/2$.
Instead, if a node needs to be removed,  choose a random node of the network remove it and rewire/remove links in order to enforce that the total number of links is the integer that  better approximates  $\avg{k}N/2$.
\end{itemize}

The Metropolis-Hastings algorithm for the exchangeable sparse networks with $N$ nodes is the same algorithm used in Ref. \cite{bianconi2022exchangeable} for exchangeable networks with finite size $N$ and is indicated below.
\begin{itemize}
\item[(1)] Start with a network of $N$ nodes having exactly $L=\avg{k}N/2$ links and in which the minimum degree is greater of equal to $\hat{m}$ and the maximum degree is smaller or equal to $K$.
\item[(2)]Iterate the following steps until equilibration:
\begin{itemize}
\item[(i)] Let ${\bf a}$ be the adjacency matrix of the network;
\item[(i)] Choose randomly a random link $\ell=(i,j)$ between node $i$ and $j$ and choose a pair of random nodes $(i',j')$ not connected by a link. 
\item[(ii)] Let ${\bf a'}$ be the adjacency matrix of the network in which the link $(i, j)$ is removed and the link  $(i',
j')$ is inserted instead. Draw a random number $r$ from a uniform
distribution in $[0, 1]$, i.e. $r\sim U(0, 1)$. 
If $r <\mbox{max}(1,e^{-\Delta H})$
where $\Delta H= H({\bf a}')-H({\bf a})$ and if the move does not violate the conditions
on the minimum and maximum degree of the network, replace ${\bf a}$ by ${\bf a'}$.
\end{itemize}
\end{itemize}
The Metropolis-Hastings algorithm can be used to sample the space of networks with variable number of nodes and given (stable) degree distribution (see Figure \ref{fig:2}).

\subsection{Monte Carlo generation of grand canonical network ensemble with given latent variable distribution}
A single instance of the grand canonical model with given latent variable distribution can be obtained by performing the following algorithm:
\begin{itemize}
    \item[1] Draw the network size $N$ from the $\pi(N)$ distribution;
    \item[2] Draw the latent variable $\theta_i$ of each node $i$ independently from the latent variable distribution $p(\theta)$.
    \item[3] Draw  each link $(i,j)$ of the network with  probability $p_{N}(\theta_i,\theta_j)$.
\end{itemize}

\begin{figure*}[htbp!]
		\centering
	\includegraphics[width=0.95\textwidth]{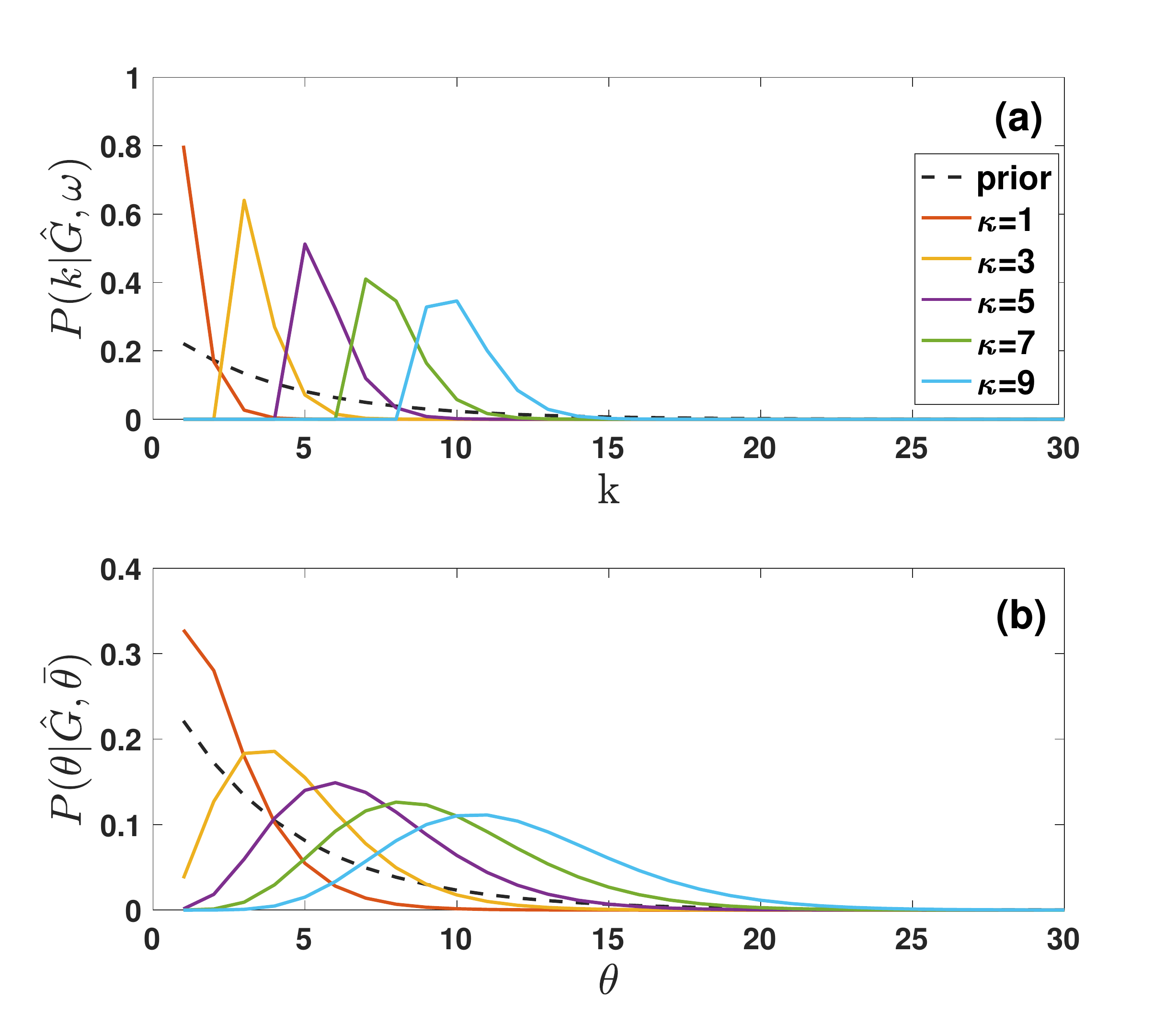}
	\caption{Marginal posterior probability for the true degree of sampled nodes (panel (a)) and for the true latent variable of sampled node (panel b). The posterior probability $P(k_i|\hat{G},\omega)$ (panel (a)) of the true degree of a sampled nodes depends on the degree $\kappa$ of the nodes in the sampled network $\hat{G}$ and is non-zero only for $k\geq \kappa$. The posterior 
probability $P(\theta|\hat{G},\bar{\theta})$ of the latent variable of a sampled node can be non-zero on the entire  range of $\theta$ values allowed by the prior. Here we have plotted $P(k_i|\hat{G},\omega)$ and $P(\theta|\hat{G},\bar{\theta})$ for different values of $\kappa$ and we have chosen $\omega=2$ and $\bar{\theta}=0.6$. The dashed lines indicate the exponential prior on the degrees (panel (a)) and on the latent variables (panel (b)).	}\label{fig:3}
	\end{figure*}
	\begin{figure*}[htbp!]
		\centering
	\includegraphics[width=0.95\textwidth]{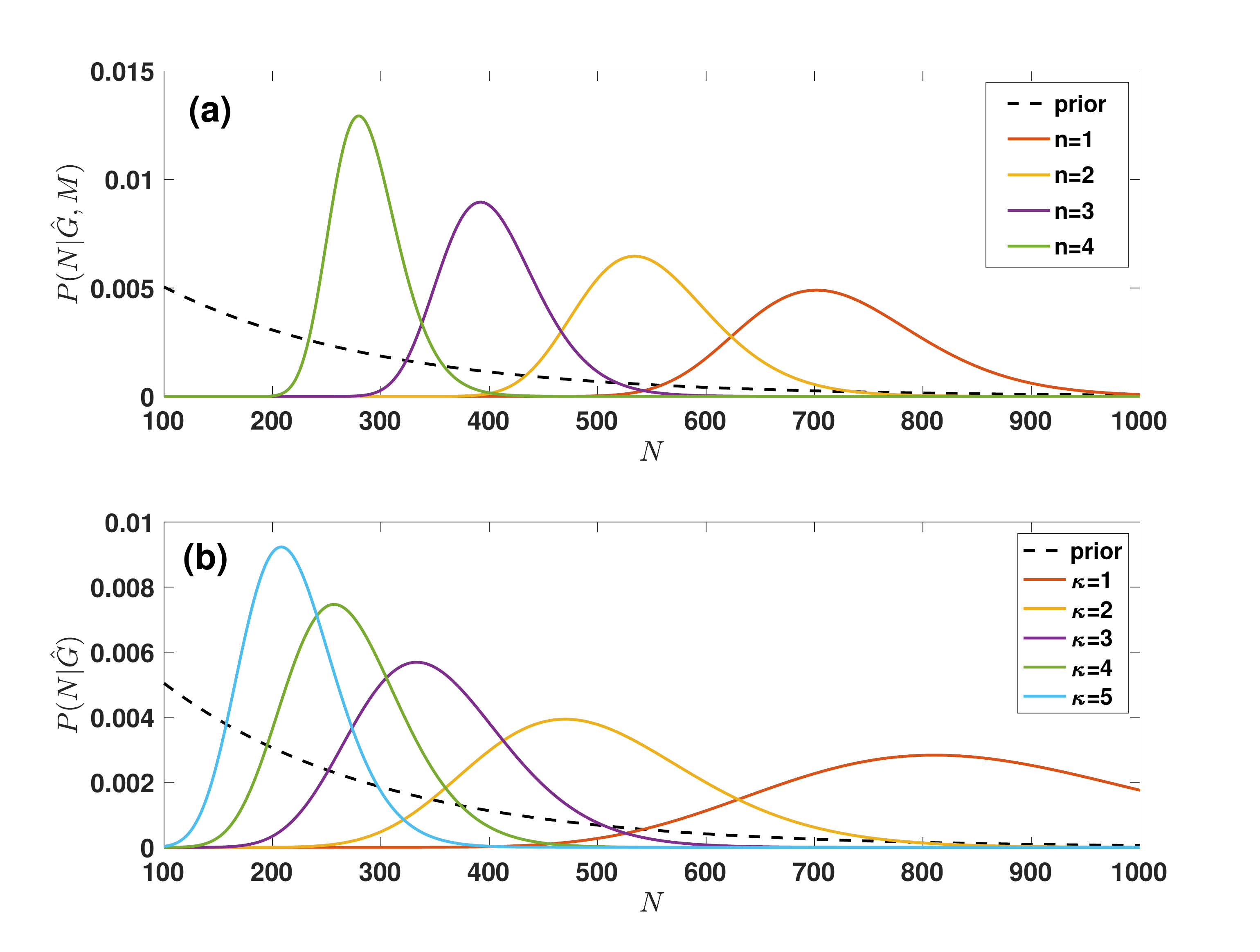}
	\caption{Marginal posterior probability for the true number of nodes in the grand canonical ensemble with given degree distribution (panel (a)) and in the grand canonical ensemble with given latent variable distribution (panel (b)). The posterior probability $P(N|\hat{G},M)$ in panel (a) of the true number of nodes depends on the total number $M$ of true but not observed links of the sampled nodes and on the total number of sampled links $\hat{L}$; the posterior probability $P(N|\hat{G})$ in panel (b) depends instead only on the degree $\kappa$ of the nodes in the sampled network $\hat{G}$.
	We took $\hat{N}=100$ and the priors given by 
	$\pi(N)\propto e^{-N/N0}$, $p(k)\propto e^{-k/m}$, $p(\theta)\propto e^{-\theta/m}$ with $N0=200$,and $m=7$.
	In panel (a) we have plotted $P(N|\hat{G},M)$ for different values of $M=(\avg{k}-n)\hat{N}$ with $n=1,2,3,4$ and $\hat{L}=\hat{N}/2$; in panel  (b) we have plotted  $P(N|\hat{G})$ assuming that $\hat{G}$ is regular with all sampled nodes having sampled degree  $\kappa=1,2,3,4,5$ . The dashed lines indicate the exponential prior $\pi(N)$ on the number of nodes.
}
	\label{fig:4}
	\end{figure*}
\section{Bayesian estimation of the network parameters  given partial knowledge of the network}

In this  section  we will use the grand canonical network ensembles for calculating the posterior distribution of the network parameters given partial information of a network $G=(V,E)$. 
In particular let us assume that we only know the subgraph $\hat{G}(\hat{V},\hat{E})$ induced by a set of nodes $\hat{V}\subset V$  of $\hat{N}=|\hat{V}|$ nodes and  of adjacency matrix ${\bf \hat{a}}$ and we do not have access to the full network $G$ with adjacency matrix ${\bf a}$. Without loss of generality let us label the nodes of the network in such a  way  that the labels $i$ with $1\leq i\leq \hat{N}$ indicate the nodes in $\tilde{V}$ (denote as  sampled nodes) and the labels $i$ with $i>\hat{N}$ indicate the nodes in $V\setminus \hat{V}$ (denoted also as unsampled or unknown nodes). We indicate with $\bm\kappa$ the  degree sequence  of the sampled network $\hat{G}$. 
Our goal is to  make a Bayesian estimation of the network size $N$ and the true network parameters given the observed subgraph $\hat{G}$.
These a posteriori estimates of the true parameters of the network can then be used to reconstruct the unknown part of the network $G$.

\subsection{Inferring the true parameters with the grand canonical ensemble with given degree distribution}
In this paragraph we will use the grand canonical ensemble with given degree distribution to find the  posterior probability distribution of the network parameters.
For convenience we will indicate with   $k_i$ the true degree of  the  sampled nodes $1\leq i\leq \hat{N}$ and we will indicate  $q_i$ the true degree of the remaining unsampled $N-\hat{N}$ nodes $\hat{N}+1\leq i\leq N$. To this end, using the Bayes rule we get the following expression for the posterior distribution of the network parameters given the observed subgraph $\hat{G}$
\bea
P(N,{\bf k},{\bf q}|{\hat{G}})=\frac{P(N)P({\bf k},{\bf q}|N){P}(\hat{G}|{\bf k},{\bf q},N)}{P(\hat{G})}
\label{Bayes}
\eea
 where 
\bea
P(N)&=&\pi(N),\nonumber \\
P({\bf k},{\bf q}|N)&=&\prod_{i=1}^{\hat{N}}p(k_i)\prod_{i=1+\hat{N}}^N p(q_i),\nonumber \\
P(\hat{G}|{\bf k},{\bf q},N)&=&e^{-\Delta_N\Sigma({\bf k,q}|\bm\kappa)},
\eea
with  $\Delta_N\Sigma({\bf k,q}|{\bm \kappa})$ given by 
\bea
\Delta_N\Sigma({\bf k,q}|{\bm\kappa})=\Sigma_N({\bf k,q})-\hat{\Sigma}_N({\bf k,q}|\bm\kappa).
\eea
Here $\Sigma_N{\bf k,q})$ indicates the entropy of the network fo size $N$ with degree sequence $[{\bf k},{\bf q}]$ whose expression is given by  the Bender-Canfield formula \cite{bender,bianconi2008entropies,anand2009entropy,anand2010gibbs} (Eq.(\ref{S})) which reads in this case
\bea
\Sigma_N({\bf k,q})=(2L)!!\left[\prod_{i=1}^{\hat{N}}k_i! \prod_{i=1+\hat{N}}^N q_i!\right]^{-1}.\eea 
Moreover $\hat{\Sigma}_N({\bf k,q}|{\bm{\kappa}})$ indicates the logarithm of the number of networks of $N$ nodes having $\hat{G}$ (with adjacency matrix ${\bf \hat{a}}$ and degree sequence $\bm\kappa$) as induced subgraph between the $\hat{N}$ sampled nodes.

Moreover in Eq. (\ref{Bayes}) $P(\hat{G})$ indicates the evidence of the data given by
\bea
P({\hat{G}})=\sum_{N}\sum_{{\bf k,q}} \pi(N)\prod_{i=1}^{\hat{N}}p(k_i)\prod_{i=1+\hat{N}}^Np(q_i)e^{-\Delta_N \Sigma({\bf k,q}|\bm\kappa)}.
\eea

Calculating the entropy $\hat{\Sigma}_N({\bf k,q}|{\bm{\kappa}})$ using  statistical mechanics methods including the use of a  functional order parameter (see Appendix), we derive the following  expression:
\bea
\hat{\Sigma}_N({\bf k,q}|{\bm\kappa})&=&\ln \left[\frac{M!(Q-M)!!}{\prod_{i=1}^{\hat{N}}(k_i-\kappa_i)!\prod_{i=1+\hat{N}}^Nq_i!}\left(\begin{array}{c}Q\\ M\end{array}\right)\delta(Q+M,2L+2\hat{L})\right]
\label{hSigma}
\eea
where $M$ indicates the number of links between the sampled nodes and the unsampled nodes and $Q$ indicates the sum over all the degrees of the unsampled nodes, i.e. 
\bea
M&=&\sum_{i=1}^{\hat{N}}(k_i-\kappa_i),\nonumber\\
Q&=&\sum_{i=1+\hat{N}}^Nq_i,
\label{MQ}
\eea
where $M$ and $Q$ need to satisfy   the constraint enforcing that the total number of true links is given by $L=\avg{k}N/2$.
Therefore, indicating with $\hat{L}=\sum_{i=1}^{\hat{N}}\kappa_i/2$,  we must impose
\bea
{Q}+M=2L-2\hat{L}.
\eea

The expression obtained for the entropy $\hat{\Sigma}_N({\bf k},{\bf q}|\bm\kappa)$ implies that the asymptotic expression for the number of networks with $N$ nodes, degree sequence $[{\bf k}, {\bf q}]$ having $\hat{G}$ as a subgraph is given by (see Appendix for the derivation) 
\bea
\mathcal{N}({\bf k},{\bf q}|\bm\kappa,N)=e^{\hat{\Sigma}_N({\bf k},{\bf q}|\bm\kappa)}=\frac{M!(Q-M)!!}{\prod_{i=1}^{\hat{N}}(k_i-\kappa_i)!\prod_{i=1+\hat{N}}^Nq_i!}\left(\begin{array}{c}Q\\ M\end{array}\right)\delta(Q+M,2L+2\hat{L}).
\eea
This expression admits a simple combinatorial interpretation. In fact the  networks with degree sequence $[{\bf k},{\bf q}]$ having as subgraph $\hat{G}$ can be constructed by adding (unsampled) links to the graph $\hat{G}$. The unsampled part of the network can be constructed by assigning to each node $i$ with $1\leq i\leq \hat{N}$ a number of stubs given by $k_i-\kappa_i$ and to each node $i$ with $i>\hat{N}$ a number of stubs given by $q_i$. The unsampled networks can then be obtained by matching the stubs pairwise with the constrains that the stubs of the first $\hat{N}$ nodes can be only matched with the stubs of the unsampled nodes $i>\hat{N}$.
Therefore the reconstructed part of the network is formed by a bipartite network between the sampled and the unsample nodes with a number of links given by $M$ and a simple network among the unsampled nodes with number of links given by $(Q-M)/2$.
The number of matchings of the $M$ links of the bipartite network is given by $M!$ the number of matching of the stubs of the simple network among unsampled nodes is $(Q-M)!!$. In order to get the number of distinct networks $G$ with degree sequence $[{\bf k},{\bf q}]$ having as subgraph $\hat{G}$ we need to divide by the number of permutations of the stubs belonging to the same nodes and we need to multiply by $Q$ choose $M$ indicating the number of ways in which we can choose the  $M$ stubs  of the unsampled nodes to be matched with the stubs of the sampled nodes. 

Given the expression for $\hat{\Sigma}_N({\bf k,q}|{\bm\kappa})$ provided by Eq.(\ref{hSigma}), we can deduce the explicit expression for $\Delta_N{\Sigma}({\bf k,q}|{\bm\kappa})$: 
\bea
\Delta_N{\Sigma}({\bf k,q}|{\bm\kappa})&=&\ln \left[\prod_{i=1}^{\hat{N}}\frac{k_i!}{(k_i-\kappa_i)!}\frac{M!(Q-M)!!}{(\avg{k}N)!!}\left(\begin{array}{c}Q\\ M\end{array}\right)\delta(Q+M,2L-2\hat{L})\right].
\label{DS}
\eea
It follows that the describe Bayesian inference assigns a probability to the model parameters a probability
\bea
{P(N,{\bf k},{\bf q}|{\hat{G}})}\propto\pi(N)\prod_{i=1}^{\hat{N}}p(k_i)\prod_{i=1+\hat{N}+1}^Np(q_i)e^{-\Delta_N{\Sigma}({\bf k,q}|{\bm\kappa})},
\label{post1}
\eea
with $\Delta_N{\Sigma}({\bf k,q}|{\bm\kappa})$ given by Eq. (\ref{DS}).
From this expression, imposing with a delta function that $M=\sum_{i=1}^{\hat{N}}(k_i-\kappa_i)$, expressing the delta in integral form  and using the saddle point to evaluate the integral, we can calculate the marginal probability $P(k_i|{\hat{G}},\omega)$ that a sampled node $i$ with $1\leq i\leq \hat{N}$ has true degree $k_i\geq \kappa_i$ given $M$  and $Q$, i.e.
\bea
P(k_i|{\hat{G}},\omega)\propto p(k_i)\frac{k_i!}{(k_i-\kappa_i)!}e^{-\omega k_i}\hat{\theta}(k_i-\kappa_i)
\eea
where $\omega$ is related to $M$ by
\bea
{M}=\sum_{i=1}^{\hat{N}}\frac{\sum_{k}p(k)k\frac{k!}{(k-\kappa_i)!}e^{-\omega k}}{\sum_{k'}p(k')\frac{k'!}{(k'-\kappa_i)!}e^{-\omega k'}}.
\eea
In Figure $\ref{fig:3}$ we show the difference between an exponential prior distribution  $p(k)$ on the degree of the nodes and the posterior marginal probability of the true degree of the sampled nodes $P(k|{\hat{G}},\omega)$ plotted for different values of  the sampled degree $\kappa$ of the same node.
Finally, we can calculate the a posteriori probability $P(N|\hat{G},M)$ that the  real networks has $N$ nodes, conditioned to $M$ and to the sampled subrgraph $\hat{G}$.
To this end we sum Eq. (\ref{post1}) over all the possible values of the degrees ${\bf k}$ and ${\bf q}$ such that Eqs.(\ref{MQ}) are satisfied.
Therefore, by inserting Eq. (\ref{DS})
into Eq.(\ref{post1}), enforcing Eqs.(\ref{MQ}) with Kronecker deltas and integrating over all the possible values of ${\bf k}$ and ${\bf q}$ we get
\bea
P(N|\hat{G},M)\propto\pi(N)\theta(N-\hat{N})C_{M,N}I^{(k)}(M)I^{(q)}(M,N),
\eea
where 
\bea
C_{M,N}&=&\frac{M!(Q-M)!!}{(\avg{k}N)!!}\left(\begin{array}{c}Q\\ M\end{array}\right),\nonumber\\
I^{(k)}(M)&=&\sum_{{\bf k}}\left[\prod_{i=1}^{\hat{N}}p(k_i)\frac{k_i!}{(k_i-\kappa_i)!}\delta\left(M,\sum_{i=1}^{\hat{N}}k_i\right)\right],\nonumber \\
I^{(q)}(M,N)&=&\sum_{{\bf q}}\left[\prod_{i=1+\hat{N}}^{{N}}p(q_i)\delta\left(Q,\sum_{i=1+\hat{N}}^{{N}}q_i\right)\right],
\eea
where $Q=\avg{k}N-2\hat{L}-M$. By expressing the Kronecker deltas in an integral form according to the expression
\bea
\delta(x,y)=\frac{1}{2\pi}\int_{-\pi}^{\pi}e^{\textrm{i}\omega(x-y)},
\eea
performing a Wick rotation and evaluating the integrals at the saddle point, we can express $I^{(k)}(M)$ and $I^{(q)}(M,N)$ as 
\bea
I^{(k)}(M)&=&\frac{1}{{2\pi}}\left[\prod_{i=1}^{\hat{N}}\sum_{k>\kappa_i}p(k)\frac{k!}{(k-\kappa_i)!}e^{-\omega^{\star} k}\right]e^{\omega^{\star}M},\nonumber \\
I^{(q)}(M,N)&=&\frac{1}{{2\pi}}\left[\sum_qp(q)e^{-\bar{\omega}^{\star}q}\right]^{N-\hat{N}}e^{\bar{\omega}^{\star}Q},
\eea
with $\omega^{\star}$ and $\bar{\omega}^{\star}$ fixed by the saddle point equations
\bea
M&=&\sum_{i=1}^{\hat{N}}\frac{\sum_{k>\kappa_i}p(k)\frac{k!}{(k-\kappa_i)!}k\  e^{-\omega^{\star} k}}{\sum_{k>\kappa_i}p(k)\frac{k!}{(k-\kappa_i)!}e^{-\omega^{\star} k}},\nonumber \\
Q&=&(N-\hat{N})\frac{\sum_{q}p(q)q\  e^{-\bar\omega^{\star} q}}{\sum_{q}p(q)e^{-\bar\omega^{\star} q}}. 
\eea
In Figure $\ref{fig:4}$ we display the marginal a posteriori distribution $P(N|\hat{G},M)$  as  function of $M$ demonstrating that the sampled network can modify significantly the prior assumptions on the total number of nodes in the network.
\subsection{Inferring the true parameters with the grand canonical ensemble with given latent variable distribution}
In this section we treat the problem of Bayesian estimation of the parameters of the true network $G$  given the sampled network $\hat{G}$ using the grand canonical model with given latent variable distribution. Let us indicate with  $\theta_i$ the latent variables of the sampled nodes $1\leq i\leq \hat{N}$ and with $\phi_i$ the latent variables of the unsampled nodes $i>\hat{N}$. Using Bayes rule we have
\bea
P(N,{\bm\theta},{\bm \phi}|{\hat{G}})=\frac{P(N)P({\bm\theta},{\bm\phi}|N){P}(\hat{G}|\bm\theta,\bm\phi,N)}{P(\hat{G})},
\label{BayesL}
\eea
where ${P}(\hat{G}|\bm\theta,\bm\phi,N)$ is independent of $\bm\phi$, i.e. ${P}(\hat{G}|\bm\theta,\bm\phi,N)={P}(\hat{G}|\bm\theta,N)$ and where
\bea
P(N)&=&\pi(N),\nonumber \\
P({\bm\theta},{\bm\phi}|N)&=&\prod_{i=1}^{\hat{N}}p(\theta_i)\prod_{i=1+\hat{N}}^N p(\phi_i),\nonumber \\
{P}(\hat{G}|\bm\theta,N)&=&\prod_{i<j|i,j\in \hat{V}}p_N(\theta_i,\theta_j)^{\hat{a}_{ij}}(1-p_N(\theta_i,\theta_j))^{1-\hat{a}_{ij}}
\eea
with $p_N(\theta_i,\theta_j)$ given by Eq. (\ref{ptheta2}) and with ${\bf \hat{a}}$ indicating the adjacency matrix of the sampled subgraph $\hat{G}$.
In Eq. (\ref{BayesL}) $P(\hat{G})$ indicates the evidence of the data given by
\bea
P({\hat{G}})=\sum_{N}\sum_{\bm\theta} \pi(N)\prod_{i=1}^{\hat{N}}p(\theta_i){P}(\hat{G}|\bm\theta,N).
\eea
Since, as we have observed previously, ${P}(\hat{G}|\bm\theta,\bm\phi,N)$ is independent of $\bm\phi$  the Bayesian estimation of the parameters $\bm\phi$ reduces simply to the prior in this case.
Therefore we focus here only on the Bayesian estimate of the latent variables $\bm\theta$, i.e.  we consider 
\bea
P(N,{\bm\theta}|{\hat{G}})=\frac{P(N)P({\bm\theta}|N){P}(\hat{G}|\bm\theta,N)}{P(\hat{G})},
\eea
with $P(N),{P}(\hat{G}|\bm\theta,N),P(\hat{G})$ having the same definition as above and 
\bea
P({\bm\theta}|N)=\prod_{i=1}^{\hat{N}}p(\theta_i).
\eea
Using the explicit expression of $p_N(\theta_i,\theta_j)$ given by Eq. (\ref{ptheta2}), we can express the likelihood ${P}(\hat{G}|\bm\theta,N)$ of the sampled network as 
\bea
{P}(\hat{G}|\bm\theta,N)
&=&\prod_{i=1}^{\hat{N}}{\theta_i^{\kappa_i}}\prod_{i<j|i,j\in \hat{V}}\left(1+\frac{\theta_i\theta_j}{N}\right)^{-1}\frac{1}{N^{\hat{L}}},
\eea
where $\hat{L}$ is the number of links of the sampled network $\hat{G}$.
In the limit $N\gg 1$ we can approximate this expression as 
\bea
{P}(\hat{G}|\bm\theta,N)&\simeq &\int d\bar{\theta}\ \prod_{i=1}^{\hat{N}}\left[{\theta_i^{\kappa_i}}e^{-\theta_i\bar{\theta}/2}\right]\frac{1}{{N}^{\hat{L}}}{\delta}\left(\bar{\theta},\sum_{j=1}^{\hat{N}}\theta_j/N\right)
\eea
With this approximation we get that the posterior probability $P(N,\bm\theta|\hat{G})$ is given by 
\bea
P(N,\bm\theta|\hat{ G})\propto&\pi(N)\frac{1}{N^{\hat{L}}}\int d\bar{\theta}\ \prod_{i=1}^{\hat{N}}\left[p(\theta_i){\theta_i^{\kappa_i}}e^{-\theta_i\bar{\theta}/2}\right]{\delta}\left(\bar{\theta},\sum_{j=1}^{\hat{N}}\theta_j/{N}\right).
\label{post2}
\eea
Calculating the marginal posterior probability of a single latent variable conditional of $\bar{\theta}$ we get
\bea
P(\theta_i|{ \hat{G}},\bar{\theta})=p(\theta_i){\theta_i^{\kappa_i}}e^{-\theta_i\bar{\theta}/2}.
\eea
In Figure $\ref{fig:3}$ we show the difference between an exponential prior distribution  $p(\theta)$ on the latent variables  of the nodes and the posterior marginal probability of the true latent variables  of the sampled nodes $P(\theta|{\hat{G}},\bar{\theta})$ plotted for different values of  the sampled degree $\kappa$ of the same node.

Stating from Eq. (\ref{post2}) we can also calculate the posterior distribution  $P(N|\hat{G})$ of the true number of nodes $N>\hat{N}$.
To this end we express the delta function in an integral form and we sum over all possible latent variables $\bm\theta$, obtaining 
\bea
P(N|\hat{G})\propto \pi(N)\theta(N-\hat{N})\frac{1}{N^{\hat{L}-1}}\frac{1}{2\pi}\int d\bar\theta d\omega e^{\textrm{i}N\omega\bar\theta}
I^{(\theta)}(\omega,\bar\theta)
\label{post3}
\eea
where $I^{(\theta)}(\omega,\bar\theta)$ is given by
\bea
I^{(\theta)}=\prod_{i=1}^{\hat{N}}\left(\sum_{\theta}p(\theta)\theta^{\kappa_i}e^{-\theta(\bar\theta/2-\textrm{i}\omega)}\right).
\eea
The integrals in Eq. (\ref{post3}) can be calculated at the saddle point getting 
\bea
P(N|\hat{G})\propto \pi(N)\theta(N-\hat{N})\frac{1}{N^{\hat{L}-1}}e^{N\frac{(\bar{\theta}^{\star})^2}{2}}\left[\prod_{i=1}^{\hat{N}}\left(\sum_{\theta}p(\theta)\theta^{\kappa_i}e^{-\theta\bar\theta^{\star}}\right)\right]
\eea
where 
\bea
\bar{\theta}^{\star}=\frac{1}{N}\sum_{i=1}^{\hat{N}}\frac{\sum_{\theta}p(\theta)\theta^{\kappa_i+1}e^{-\theta\bar\theta^{\star}}}{\sum_{\theta}p(\theta)\theta^{\kappa_i}e^{-\theta\bar\theta^{\star}}}.
\eea
In Figure $\ref{fig:4}$ we display the marginal  a posteriori distribution $P(N|\hat{G})$ on the true number of nodes  in the simplified assumption in which $\hat{G}$ is regular and all degree $\kappa$ are the same demonstrating that the sampled network can modify significantly the prior assumptions on the total number of nodes in the network.

\section{Conclusions}
In this paper we have proposed grand canonical network ensembles formed by networks of varying  number of nodes.
The grand canonical network ensembles we have introduced are both sparse and exchangeable, i.e. have a finite average degree and are invariant under permutation of the node labels.
The grand canonical ensembles are hierarchical network models in which first the network size is selected, then the degree sequence (or the sequence of latent variables) and finally the  network adjacency matrix is selected.
The model circumvents the difficulties imposed by  the Aldous-Hoover theorem that states that exchangeable infinite sparse network ensembles vanish, as the network is a mixture of finite networks, although the networks can have an arbitrarily large network size. 
Here we show how the grand-canonical ensembles can be used to perform a  Bayesian estimation of the network parameters when only partial information about the network structures is known. This a posteriori estimation of the network parameters can then be used for network reconstruction.

The grand canonical framework for sparse exchangeable network ensembles  is here described for the case simple networks but has the potential to be extended to generalized network structures including directed, bipartite networks, multiplex networks and simplicial complexes following the lines outlined in Ref.\cite{bianconi2022exchangeable}.

In conclusion we hope that this work, proposing hierarchical grand canonical network ensembles able to treat networks of different size and  relating network theory to statistical mechanics will stimulate further results of mathematicians, physicists, and computer scientists  working in network science and related  machine learning problems.

\bibliography{exchangeable_bib}
%\end{document}
\appendix
\section*{Appendix: Derivation of $\hat{\Sigma}_N({\bf k,q}|\bm\kappa)$}
%\label{Ap}
In this Appendix our goal is to derive the asymptotic expression of $\hat{\Sigma}_N({\bf k,q}|\bm\kappa)$ in the limit of large network size of the sampled network $\hat{N}\gg 1$, and of the true network $N=(1+\alpha)\hat{N}\gg 1$ with $\alpha>0$.

Let us assume that the sampled subgraph $G$ is the network between the sampled nodes $1\leq i\leq \hat{N}$ and has adjacency matrix $\hat{\bf a}$.
The true network is instead formed by $N$ nodes with adjacency matrix ${\bf a}$. We assume that ${\bf a}$ has the block structure given by 
\bea
{\bf a}=\left(\begin{array}{cc} {\bf \hat{a}}& {\bf b}\\ {\bf b}^{\top}&{\bf \tilde{a}}\end{array}\right),
\eea where ${\bf b}$ indicates  the $\hat{N}\times \alpha\hat{N}$ matrix  between sampled nodes and the unsampled node and ${\tilde{\bf a}}$ indicates tha $(\alpha\hat{N})\times(\alpha\hat{N})$ adjacency matrix among the unsampled nodes.
As we have mentioned in the main text $\hat{\Sigma}_N({\bf k,q}|\bm\kappa)$ is the logarithm of the number $\mathcal{N}({\bf k,q}|\bm\kappa,N)$ of networks (or adjacency matrices ${\bf a}$)  with degree sequence $[{\bf k,q}]$ and admitting as a subgraph $\hat{G}$ having sampled degree sequence $\bm\kappa$.
In statistical mechanics we also call $\mathcal{N}({\bf k,q}|\bm\kappa,N)$ the partition function of its corresponding statistical mechanics network model, and we indicate it by $Z$. 
In terms of the matrices ${\bf b}$ and ${\bf \tilde a}$ the partition function $Z=\mathcal{N}({\bf k,q}|\bm\kappa)=\exp\left(\hat{\Sigma}_N({\bf k,q}|\bm\kappa)\right)$
can be written as 
\bea
Z&=&\sum_{{\bf b},{\bf \tilde{a}}}\prod_{i=1}^{\hat{N}}\delta\left(k_i-\sum_{j=1}^N a_{ij}\right)\prod_{i=1+\hat{N}}^{N}\delta\left(q_i-\sum_{j=1}^N a_{ij}\right)\delta\left(2L-\sum_{i=1}^Nk_i\right)\eea
Expressing the Kronecker deltas in the integral form and performing the sum over the elements of the matrices ${\bf b}$ and ${\bf \tilde a}$ we obtain
\bea
Z&=&\int {{\mathcal D}\omega}\int {{\mathcal D}\tilde\omega}\int \frac{d\lambda}{2\pi} e^{G(\bm{\omega},\tilde{\bm\omega},\lambda)}
\label{Zs}
\eea
with 
%\begin{widetext}
\bea
G(\bm{\omega},\tilde{\bm\omega},\lambda)&=&\sum_{i=1}^{\hat{N}} [\mi \omega_i (k_i-\kappa_i)]+\sum_{i=1+\hat{N}}^N [\mi\tilde{\omega}_iq_i]
+\sum_{i=1}^{\hat{N}}\sum_{j=1}^{\hat{N}}\ln (1+e^{-\mi\omega_i-\mi\tilde{\omega}_j-\rm{i}\lambda})\nonumber \\
&+&\frac{1}{2}\sum_{i=\hat{N}+1}^{N}\sum_{j=\hat{N}+1}^{N}\ln (1+e^{-\mi\tilde{\omega}_i-\mi\tilde{\omega}_j-\rm{i}\lambda})+\mi \lambda (L-\hat{L}),
\eea
and with ${\mathcal D}\omega=\prod_{i=1}^{\hat{N}} [d\omega_i/(2\pi)]$ and  ${\mathcal D}\tilde\omega=\prod_{i=1+\hat{N}}^{{N}} [d\tilde\omega_i/(2\pi)]$. 
Let us now introduce the functional order parameters \cite{courtney2016generalized,bianconi2008entropies,monasson1997statistical}
\bea
c_{\kappa,k}(\omega)&=&\frac{1}{\hat{N}\hat{P}({\kappa,k})}\sum_{i=1}^{\hat{N}}\delta(\omega-\omega_i)\delta(k,k_i)\delta(\kappa,\kappa_i),\nonumber \\
\rho_q(\tilde{\omega})&=&\frac{1}{\alpha\hat{N}\tilde{P}(q)}\sum_{i=1+\hat{N}}^N\delta(\tilde{\omega}-\tilde{\omega}_i)\delta(q,q_i),
\eea
where  $\hat{P}(k,\kappa)$ is the fraction of sampled nodes with degree $\kappa$ in the sampled network and total inferred degree $k$; $\tilde{P}(q)$ is the fraction of unsampled nodes with  degree $q$. Moreover we have indicated with $L=\avg{k}N/2$ and with $\hat{L}=\sum_{i=1}^{\hat{N}}\kappa_i/2$.
By enforcing the definition of the order parameters with a series of delta functions we obtain
\bea
&&1=\int dc_{\kappa,k}(\omega)\delta\left(c_{\kappa,k}(\omega)-\frac{1}{\hat{N}\hat{P}({\kappa,k})}\sum_{i=1}^{\hat{N}}\delta(\omega-\omega_i)\delta(k,k_i)\delta(\kappa,\kappa_i)\right)\nonumber \\
&&= \int \frac{d\hat{c}_{\kappa,k}(\omega)dc_{\kappa,k}(\omega) }{2\pi/(\hat{N}\hat{P}(\kappa,k)\Delta \omega)}\exp\left[{\mi\Delta\omega \hat{c}_{\kappa,k}(\omega)[\hat{N}\hat{P}(\kappa,k)c_{\kappa,k}(\omega)-\sum_{i=1}^{\hat{N}}\delta(\omega-\omega_i)\delta(k,k_i)\delta(\kappa,\kappa_i)]}\right].\nonumber\\
&&1=\int d\rho_q(\tilde\omega)\delta\left(\rho_q(\tilde\omega)-\frac{1}{\alpha\hat{N}{\tilde{P}(q)}}\sum_{i=1+\hat{N}}^N\delta(\tilde\omega-\tilde\omega_i)\delta(q,q_i)\right)\nonumber \\
&&= \int \frac{d\hat{\rho}_q(\tilde\omega)d\rho_q(\tilde\omega) }{2\pi/(\alpha\hat{N}\tilde{P}(q)\Delta \tilde\omega)}\exp\left[{\mi\Delta\tilde\omega \hat{\rho}_q(\tilde\omega)[\alpha\hat{N}\tilde{P}(q)\rho_q(\tilde\omega)-\sum_{i=1+\hat{N}}^{{N}}\delta(\tilde\omega-\tilde\omega_i)\delta(q,q_i)]}\right].\nonumber
\label{deltasc}
\eea

After inserting these expressions into the partition function  in the limit $\Delta\omega\to 0$, indicating with $\sum^{\prime}$ the sum over the allowed degree range  we obtain
\bea
Z=\sum_{\bm\kappa}^{\prime}\sum_{\bf k}^{\prime}\sum_{\bf q}^{\prime}\int \prod_{\kappa,k}{\mathcal D}{c}_{\kappa,k}({\omega})
\int \prod_{\kappa,k}{\mathcal D}\hat{c}_{\kappa,k}({\omega})
\int \prod_{q}{\mathcal D}{\rho}_q (\tilde{\omega})\int \prod_q {\mathcal D}\hat{\rho}_q(\tilde{\omega}) \int \frac{d\lambda}{2\pi} 
e^{\hat{N}f}\nonumber
\eea
with $f=f(c({\omega},k),\hat{c}(\omega, k),\rho(\tilde\omega,q),\hat\rho(\tilde\omega,q),\lambda,h)$
given by 

\bea
&&f=\sum_{\hat{m}\leq \kappa\leq K}\sum_{\kappa\leq k\leq K}\hat{P}(\kappa,k)\mi\int d\omega \hat{c}_{\kappa,k}(\omega)c_{\kappa,k}(\omega)+\alpha\mi\int d\omega \sum_{\hat{m}\leq q\leq K}\tilde{P}(q)\hat{\rho}_q(\tilde\omega)\rho_q(\tilde\omega)\nonumber \\
&&+\mi\lambda(L-\hat{L})/\hat{N}
+\Psi+\sum_{\hat{m}\leq \kappa\leq K} \sum_{\kappa \leq k\leq K}\hat{P}(\kappa,k) \ln \int \frac{d\omega}{2\pi} e^{\mi\omega (k-\kappa)-\mi\hat{c}_{\kappa,k}(\omega,k)}\nonumber \\
&&+\alpha \sum_{\hat{m}\leq q\leq K}\tilde{P}(q)\ln \int \frac{d\tilde\omega}{2\pi}  e^{\mi\tilde\omega q-\mi\hat{\rho}_q(\tilde\omega)},
\label{fsimple}
\eea
where $\Psi$ is given by 
\bea
\Psi=\frac{\alpha^2\hat{N}}{2} \sum_{\hat{m}\leq q\leq K,\hat{m}\leq q'\leq K} \tilde{P}(q)\tilde{P}(q')\int d\omega \int d\tilde\omega'\rho_q(\tilde\omega) \rho_{q'}(\tilde\omega')\ln \left(1+e^{-\mi\tilde\omega-\mi\tilde\omega'-\mi\lambda}\right)\nonumber\\
+\alpha\hat{N}\sum_{\hat{m}\leq \kappa\leq K}\sum_{\kappa \leq k\leq K}\hat{P}(\kappa,k) \sum_{\hat{m}\leq q\leq K}\tilde{P}(q) \int d\omega \int d\tilde\omega  c_{\kappa,k}(\omega) \rho_q(\tilde\omega)\ln \left(1+e^{-\mi\omega-\mi\tilde\omega-\mi\lambda}\right),\nonumber \label{Psi}
\eea
and where the functional measures are defined as \bea {\mathcal D}c_{\kappa,k}(\omega)&=&\lim_{\Delta \omega\to 0}\prod_{\omega}[dc_{k\kappa}(\omega) \sqrt{\hat{N}\hat{P}(\kappa,k)\Delta\omega/(2\pi)}]\nonumber \\ {\mathcal D}\hat{c}_{\kappa,k}(\omega)&=&\lim_{\Delta \omega\to 0}\prod_{\omega}[d\hat{c}_{\kappa,k}(\omega) \sqrt{\hat{N}\hat{P}(\kappa,k)\Delta\omega/(2\pi)}],\nonumber \\ {\mathcal D}\rho_{q}(\tilde\omega)&=&\lim_{\Delta \tilde\omega\to 0}\prod_{\tilde\omega}[d\rho_{q}(\tilde\omega) \sqrt{\hat{N}\alpha\tilde{P(q)}\Delta\tilde\omega/(2\pi)}],\nonumber \\
{\mathcal D}\hat{\rho}_{q}(\tilde\omega)&=&\lim_{\Delta \tilde\omega\to 0}\prod_{\tilde\omega}[d\hat{\rho}_{q}(\tilde\omega) \sqrt{\hat{N}\alpha\tilde{P}(q)\Delta\tilde\omega/(2\pi)}].
\eea
By putting
\bea
e^{-i\lambda}=\frac{z}{\hat{N}},
\eea
and performing a Wick rotation in $\lambda$ and assuming $z/\hat{N}=e^{-\mi\lambda}$ real and much smaller than one, i.e. $z/\hat{N}\ll 1$ which is allowed in the sparse regime, we can linearize the logarithm and express $\Psi$ as 
\bea
\Psi=z\alpha\nu\left(\frac{1}{2}\alpha\nu+\hat{\nu}\right),
\eea with 
\bea
{\nu}&=&\sum_{\hat{m}\leq q\leq K} \tilde{P}(q)\int d\tilde\omega \rho_q(\tilde\omega) e^{-\mi\tilde\omega}.\nonumber \\
\hat{\nu}&=&\sum_{\hat{m}\leq \kappa\leq K}\sum_{\hat{\kappa},\leq k\leq K}\hat{P}(\kappa,k)\int d\omega c_{\kappa,k}(\omega) e^{-\mi\omega}.
\eea
%\end{document}
The saddle point equations determining the value of the partition function can be obtained by performing the (functional) derivative of $f$ with respect to the functional order parameters, obtaining
\bea
-\mi\hat{c}_{\kappa,k}(\omega)&=&z\alpha\nu e^{-\mi\omega},\nonumber \\
-\mi\hat{\rho}_{q}(\omega)&=&z(\alpha\nu+\hat{\nu}) e^{-\mi\tilde{\omega}},\nonumber \\
c_{\kappa,k}(\omega)&=&\hat{P}(\kappa,k)\frac{\frac{1}{2\pi}e^{\mi\omega (k-\kappa)-\mi\hat{c}_{\kappa,k}(\omega)}}{\int \frac{d\omega'}{2\pi} e^{\mi\omega' (k-\kappa)-\mi\hat{c}_{\kappa,k}(\omega')}},\nonumber \\
\rho_q(\tilde\omega)&=&\tilde{P}(q)\frac{\frac{1}{2\pi}e^{\mi\tilde\omega q-\mi\hat{\rho}_q(\tilde\omega)}}{\int \frac{d\tilde\omega'}{2\pi}  e^{\mi\tilde\omega' q-\mi\hat{\rho}_{q}(\tilde\omega')}},\nonumber \\
2\frac{L-\hat{L}}{\hat{N}}&=&z\alpha\nu\left(\alpha\nu+2\hat{\nu}\right).
\eea
%\end{document}
Let us first calculate the integrals 
\bea
I_{\kappa,k}&=&\int \frac{d\omega}{2\pi}  e^{-\mi\omega (k-\kappa)-\mi\hat{c}_{\kappa,k}(\omega)}=\frac{1}{(k-\kappa)!}(z\alpha\nu)^{k-\kappa},\nonumber \\
I_{q}&=&\int \frac{d\tilde\omega}{2\pi}  e^{-\mi\tilde\omega q-\mi\hat{\rho}_{q}(\tilde\omega)}=\frac{1}{q!}[z(\alpha\nu+\hat{\nu})]^q,
\eea
Using these expressions for the integral we can write the functional order parameters as 
\bea
\hspace{-5mm}c_{\kappa,k}(\omega)=\hat{P}(\kappa,k)\frac{1}{2\pi}\frac{e^{\mi\omega (k-\kappa)+(z\alpha\nu)e^{-\mi\omega}}}{I_{\kappa,k}},\nonumber\\
\hspace{-5mm}\rho_{q}(\tilde\omega)=\tilde{P}(q)\frac{1}{2\pi}\frac{e^{\mi\tilde\omega q+[z\nu(\alpha\nu+\hat{\nu})]e^{-\mi\tilde\omega}}}{I_{q}}.\nonumber\\
\label{Csimple1}
\eea
%\end{document}
With this expression, using a similar procedure  we can express $\nu$ as 
\bea
\hat{\nu} &=&\int d\omega  \sum_{\hat{m}\leq \kappa\leq K}\sum_{\kappa\leq k\leq K} c_{\kappa,k}(\omega) e^{-\mi\omega}
=\sum_{\hat{\kappa}\leq k\leq K}\hat{P}(\kappa,k)(k-\kappa)(\alpha z\nu)^{-1}.\nonumber \\
{\nu} &=&\int d\tilde\omega \sum_{\hat{m}\leq q\leq K} \rho_{q}(\tilde\omega) e^{-\mi\tilde\omega}
=\sum_{\hat{m}\leq q\leq K}\tilde{P}(q)q[z(\alpha\nu+\hat{\nu})]^{-1}.
\eea
Combing these equations with the last saddle point equation 
it is immediate to show that $z,\nu$ and $\hat\nu$ are given by
\bea
z&=&1,\nonumber \\ 
\alpha\nu&=&\sqrt{(Q-M)/\hat{N}}, \nonumber \\ 
\hat{\nu}&=&\frac{M/\hat{N}}{\sqrt{(Q-M)/\hat{N}}}.
\eea
with 
\bea
2L-2\hat{L}=M+Q.
\eea
Calculating the free energy  $\hat{N}f$ at the saddle point, we get 
\bea
\hat{N}f&=&-\frac{1}{2}(Q-M)-M+(L-\hat{L})\ln \hat{N}
+\hat{N}\sum_{\hat{\m}\leq \kappa\leq K}\sum_{\hat{\kappa}\leq k\leq K}\hat{P}(\kappa,k)\ln \frac{(\alpha\nu)^{k-\kappa}}{(k-\kappa)!}\nonumber \\&&+\alpha\hat{N}\sum_{\hat{m}\leq q\leq K}\tilde{P}(q)\ln \frac{[\alpha\nu+\hat{\nu}]^{q}}{q!},
\eea
%\end{widetext}
which leads to the following asymptotic expression for $Z=\mathcal{N}({\bf k,q}|\bm\kappa,N)=\exp\left(\hat{\Sigma}_N({\bf k},{\bf q}|\bm\kappa)\right)$
\bea
Z=\mathcal{N}({\bf k,q}|\bm\kappa,N)&\simeq &\frac{M!(Q-M)!!}{\prod_{i=1}^{\hat{N}}k_i!\prod_{i=1+\hat{N}}^Nq_i!}\left(\begin{array}{c} Q\\ M\end{array}\right).
\eea

\end{document}